\documentclass{article}
\def\P{{\textbf{P}}}

\def\K{{\textbf{K}}}
\def\H{{\textbf{H}}}
\def\R{{\textbf{R}}}

\def\X{{\textbf{X}}}
\def\Y{{\textbf{Y}}}

\def\ke0{{\mathcal{KE}_0}}
\def\pe0{{\mathcal{PE}_0}}

\def\vu{\bf{u}}

\usepackage{graphicx}

\begin{document}
\begin{titlepage}
\title{Using data assimilation in laboratory experiments of 
geophysical flows}
\author{M. Galmiche, J. Sommeria, E. Thivolle-Cazat and J. Verron\\
Laboratoire des Ecoulements G\'eophysiques et Industriels \\
BP53 38041 Grenoble CEDEX 9, France
}
\maketitle
\begin{abstract}
Data assimilation is used in numerical simulations of 
laboratory experiments in a stratified, rotating fluid.  
The experiments are performed on the large Coriolis turntable (Grenoble,
France), which achieves a high degree of similarity
with the ocean, and the simulations are performed with a two-layer 
shallow water model. 
Since the flow is measured with a high level of precision and resolution, a detailed
analysis of a forecasting system is feasible. Such a task is much more difficult
to undertake at the oceanic scale because of
the paucity of observations and problems of accuracy and data sampling. This
opens the way to an
experimental test bed for operational oceanography. To illustrate this, 
some results on the baroclinic
instability of a two-layer vortex are presented.
\end{abstract}
\end{titlepage}

\section{\label{intro}Introduction : operational issues}

An increasing interest in operational oceanography has developed in 
recent years.
A number of pre-operational projects have emerged at the 
national and international scale, most of them
coordinating their activites within  the international Global Ocean
Data Assimilation
Experiment (GODAE).

The heart of operational systems consists of three main components : the
observation system, the
dynamical model and the data assimilation scheme.
Thanks to recent advances in
satellite and in-situ observations,
numerical modelling,
assimilation techniques and computer technology, the
operational systems have now acquired some degree of maturity.
However, there are still a number of issues that must be solved in 
applications, and validation tests are needed.

The ideal method for validating the overall forecasting system would be 
to compare results 
with independent
oceanic observations, i.e. observations that are not used in the
assimilation process.
However, such observations are rare because in-situ surveys are difficult
to undertake and extremely expensive, 
particularly in the deep ocean.
Another problem is that, because assimilation is only approximate, forecast errors may be due
not only to the model itself, but also  
to the temporal growth of imperfections in the initial condition. It is therefore 
difficult to objectively identify the model errors on the one hand and the 
assimilation errors on the other. 

Alternatively,
analytical solutions of simple flows with well-defined initial and boundary 
conditions can be used as a reference to unravel
some aspects of
the model error components. 
However, 
such analytical solutions are limited to some extremely
simplified flow configurations.  

In this letter, a new, experimental approach to these problems 
is presented. Laboratory experiments and numerical, shallow water simulations 
of simple oceanic flows are performed and sequential data assimilation is used 
as a tool to keep the numerical simulation close to the experimental reality. 
By contrast with real-scale oceanic measurements, the experimental measurements 
are available with a high level of precision and resolution. 
The general methodology is given  
in Section 2. In Section 3, the example of un unstable vortex 
in a two-layer, rotating fluid is presented 
as an illustration of the experimental test-bed. In particular, 
the behaviour of the model is studied when data assimilation is stopped. 
The vortex deformation and splitting as predicted by the numerical simulation is 
then compared to the real flow evolution.

\section{\label{corio}The Coriolis test-bed}

Laboratory experiments are of particular interest
as test cases for operational systems, filling the gap between the
oversimplified analytical
solutions and the full complexity of real oceans.
On the one hand, they are much more realistic than
any numerical or theoretical ``reality'', provided that the experimental
facility allows good similarity
with the ocean. On the other hand, data are available with much
better space and time resolution than  actual scale oceanic measurements.
Furthermore, a great number of experiments can be performed and compared to
one other.  
Such comparisons
are obviously impossible at the real scale because of the ever changing
flow conditions in the ocean.
The flow parameters can also be easily varied to perform
parametric studies. 

Thanks to its large size (13 meter diameter),
the Coriolis turntable (Grenoble, France) is a unique facility which enables 
oceanic flows to be reproduced with a good level of similarity (see Fig.1). 
It is possible to come close to inertial regimes, i.e. with limited effects 
of viscosity and centrifugal forces.
Various experiments can be performed on the turntable
in multi-layer stratified salt water, such as experiments on vortices or
boundary currents. 

\begin{figure}[p]
\centering
\includegraphics
[scale=0.8,bb=77 51 571 597,clip]
{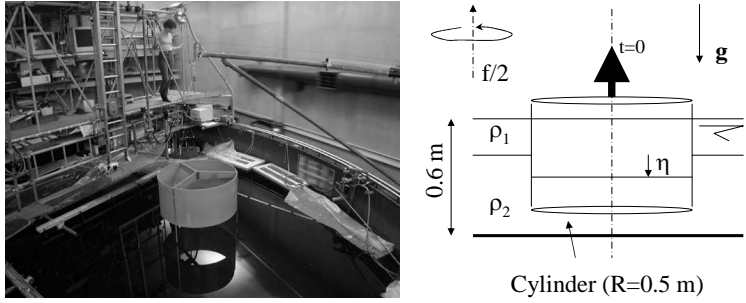}
\caption{
Picture of the Coriolis turntable  with the
setup of the two-layer vortex experiment. The layers have density $\rho_1$
and $\rho_2>\rho_1$ and
undisturbed thickness $H_1$=12.5 cm and $H_2$=50 cm.  For the experiment
presented in this paper,
the relative density
difference is  1.0 10$^{-3}$, the initial displacement of the interface is $\eta_0=-H_1$
inside the cylinder, and the tank
rotation period is 40 s. The corresponding Rossby radius of deformation is
12.5 cm.
At $t=0$ the cylinder is removed.
}
\label{image}
\end{figure}
%

Our approach relies on numerical simulation of such laboratory 
experiments using data assimilation, in 
a similar way to real-scale ocean forecasting systems. 
A major difference with the real ocean is that the measured quantity here is the 
velocity field in several horizontal planes 
instead of scalar quantities measured only at the surface or
along vertical sections. The elevation of 
the interface between the layers is not measured in the experiments. It is 
treated as an output of the asssimilation process (see
section
\ref{examp}). 

The velocity field is measured in horizontal
planes using CIV (Correlation Image Velocimetry): 
particle tracers are illuminated
by a horizontal laser sheet and a camera is used to visualize the
particle displacements from above, leading,
after numerical
treatment,
to the horizontal velocity field.
The rms measurement error in particle displacement is about 0.2 pixels, as 
determined by  Fincham \& Delerce (2000), and the 
errors found in neighboring points are not correlated. The resulting error in velocity 
about 3\% of the maximum velocity. 

In parallel with these measurements, numerical simulations are performed.  
The system is modelled as a
multi-layer fluid with hydrostatic approximation for which the  
variables are the horizontal velocity components $u(x,y,i)$ and $v(x,y,i)$,
and the layer thickness
$h(x,y,i)$, where $x$ and $y$ are the horizontal coordinates and
$i$ is the layer index. The basic shallow-water equations are solved using 
MICOM (Miami Isopycnic Coordinate
Ocean Model, Bleck \& Boudra 1986) in its simplest version. 

The measured velocity field is assimilated into the simulations at
each measurement point
using an adaptive version of the Singular Evolutive 
Extended Kalman (SEEK) filter, a method adapted for oceanographic purposes
on the basis of the Kalman filter.  
Each data assimilation provides a new dynamical state which optimally 
blends the model prediction and the measured data, in accounting for their respective error. 
The forecast state vector $\X^f$ is replaced by the analysed state vector
$\X^a=\X^f + \K [\Y^o - \H \X^f]$ ,where $\Y^o$ is the observed 
part of the state vector (i.e. the velocity field 
in the measurement domain), $\H$ is the observation operator and $\K$ is the Kalman 
gain defined by $\K=\P^f  \H^T \; [\H  \P^f  \H^T  +  \R]^{-1}$. Here, 
$\P^f$ and $\R$ are the forecast error and observation error covariance matrices respectively. 
The observation errors are here supposed to be uniform, and 
the multi-variate correlations between the variables are described as
components on Empirical
Orthogonal Functions (EOF's) computed from the model statistics, providing an estimation of $\P^f$. 
The reader is referred to Pham, Verron \&  Roubaud (1998) and 
Brasseur, Ballabrera-Poy \& Verron (1999) for
mathematical details.

\medskip

\section{\label{examp}Example : Baroclinic instability of a two-layer vortex}
Among the various  
experiments performed on the Coriolis turntable, concerning, for example, 
baroclinic instability and coastal currents,  
the study of the baroclinic instability of a two-layer vortex is presented in the present letter 
because it provides a good illustration of
the experimental test-bed. This flow problem  
is of particular interest because
simple experiments
are feasible as well as numerical simulations, although it is quite a 
complex non-linear process (e.g. Griffiths and Linden 1981) and plays a crucial role in the
variability of the real ocean. The
initial conditions are well defined and the lateral boundaries have no
significant influence. 

A cylinder of radius $R=$ 0.5 m is initially introduced in a two-layer fluid
across
the interface (see Fig. 1). A displacement $\eta_0$ of
the interface is produced inside the cylinder, and at $t=0$
the cylinder is
rapidly removed.
A radial gravity current is then initiated,
which is deviated
by the Coriolis force, resulting in the formation of a vortex in the upper
layer after damping of inertial oscillations.
A vortex of opposite sign is produced in the lower layer, and the
resulting vertical shear
is a source of baroclinic instability.
The main control parameter in this system is
$\gamma=R/R_D$, where $R_D$ is the Rossby deformation radius.
The results presented here were obtained with
$\gamma=4$. The vortex then undergoes baroclinic
instability which gives
rise to splitting into two new vortices.

The experimental vortex is dynamically similar to an oceanic vortex
with a radius of the order of $100$ km at mid-latitude (the radius of deformation
is typically 25 km). 
In the experiments, 
vortex instability
takes place in typically 20 rotation periods of the tank, corresponding to
about 30 days at mid-latitude
(taking the inverse of the Coriolis parameter 
as the relevant time unit).
The
ratio of the vertical to the horizontal scales is distorted by a factor
of 10 in the experiments.  
This is not important provided that the hydrostatic approximation is valid.


The velocity field
is  measured in each layer every 11 s, which is half the observed
period of inertial oscillations. Since we are interested in the slow
balanced dynamics, we eliminate the residual
inertial oscillations by averaging two
successive fields for data assimilation.
The velocity data obtained are assimilated in the numerical model at each grid point 
in the measurement domain (2.5 m
$\times$ 2.5 m). 
In the numerical simulations, the system is modelled as a two-layer fluid
with a standard biharmonic dissipation term
and the simulation domain is 5 m wide (i.e. twice as large as the measurement domain)  
in order to avoid
spurious confinement by boundaries. The simulations are performed using 100$^2$ 
or 200$^2$ 
grid points in each layer.

A good fit is then obtained between the model and the experimental data, as
shown in Fig. 2. The irregular
shape of the vortex, the  position of its center and the presence of residual
currents in its vicinity
are well represented. The elongation of the
vortex and the formation of two new, smaller vortices are also well reproduced.
Data assimilation provides us with an indirect measurement of the interface
depth, also shown in Fig. 2. No significant inertio-gravity wave is excited 
in the simulation after data assimilation is performed, showing that the interface 
position is well determined, without any spurious imbalance effect. 

The initial development of baroclinic instability is well described by
the growth of mode two
(calulated using a polar Fourier decomposition of the radial velocity
field along a circle of radius R). Excellent agreement
between the model and the observation
is obtained when data assimilation is performed, as shown in Fig. 3. The
growth of this mode is considerably delayed
in the model without data assimilation, as the initial perturbation is
smaller than in the experiments.

The rms
distance between the forecast and measured velocity fields is plotted in
Fig. 4.
After a few assimilation cycles, this distance remains of
the order of 0.6 mm.s$^{-1}$, close to the experimental errors 
(3\% of the maximum velocity, i.e. about 0.5 mm.s$^{-1}$). 
Similar agreement is obtained in both layers.  

The state vector obtained at a given time can be used as an initial 
condition to test the free model. 
To do so, we stop the assimilation at time $t=75s$ and measure the growth
of the rms distance between
the laboratory experiments and the free model run with this new initial
state, as shown in Fig. 4.
This growth can be due
either to the amplification of small initial errors, or to limitations of
the dynamical model. 
It is actually possible to show that the sensitivity to the initial 
condition is not the dominant effect, as observed in Fig. 5. 
It is clear in this figure that the divergence of 
the model from the experimental reality is not sensitive, over the short term, to small 
variations in the initial condition. 
The model diverges from reality on a timescale of around 3000s, 
which is about 30 times the typical advection timescale of the
flow $2R/U \simeq 100s$ (where $U \simeq 1$ cm.s$^{-1}$ is the 
order of magnitude of the velocity within the flow). 
The model error is therefore 
about $1/30$ of the dominant advective term. This error is actually small 
but seems to be systematic. 


The results are similar when 100$^{2}$ or 200$^{2}$ numerical 
grid points are used in 
each layer (see Fig. 4, 5 and 6). The effect of dissipation and friction was also 
investigated in various test runs. 
The rms distance to observations obtained in the most representative 
of these test runs is plotted in Fig. 4. 
The results show that the model errors persist 
when the numerical viscosity 
coefficient is changed or when an Ekman friction term is added in the momentum 
equation. 
We notice that, in all cases, vortex splitting occurs 
faster than in the experimental reality (Fig. 6).  
It is therefore likely that the basic simplifying 
assumptions of the hydrostatic, two-layer shallow-water formulation, rather 
than resolution, dissipation or friction problems, are responsible for the limitations 
of the model.  
For instance, the interface 
between the layers may have a finite thickness in reality, leading to effects that 
cannot be reproduced in the two-layer simulation. Also, the hydrostatic approximation may 
slightly enhance the 
growth of baroclinic instability, as shown in the theoretical study 
of non-hydrostatic effects by Stone (1971).  

In the last stage of our testing procedure, we perform assimilation
using only upper layer data and check how the behavior of the
lower layer is reconstructed. The results are shown in Fig. 6.  
Although some local discrepancies 
are observed in the bottom layer 
compared to the measured velocity field, 
the global flow field is well reproduced.  

\begin{figure}[p]
\centering
\includegraphics
[scale=0.8,bb=77 51 571 597,clip]
{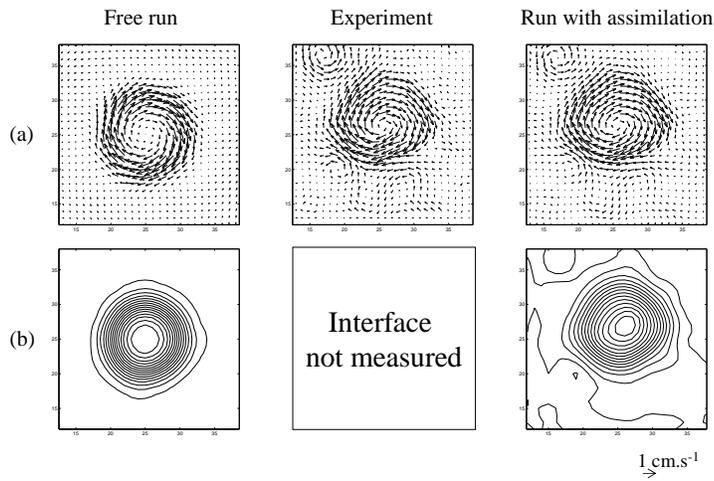}
\caption{
Velocity field in the top layer (a) and interface depth (b) at $t=75s$ in
the free run,
in the experiment and
in the simulation performed with data assimilation every 22s. For clarity,
only 25$^2$ vectors are plotted.
The rms measurement error is about 0.5 mm.s$^{-1}$.
}
\end{figure}

\begin{figure}[p]
\centering
\includegraphics
[scale=0.8,bb=77 51 571 597,clip]
{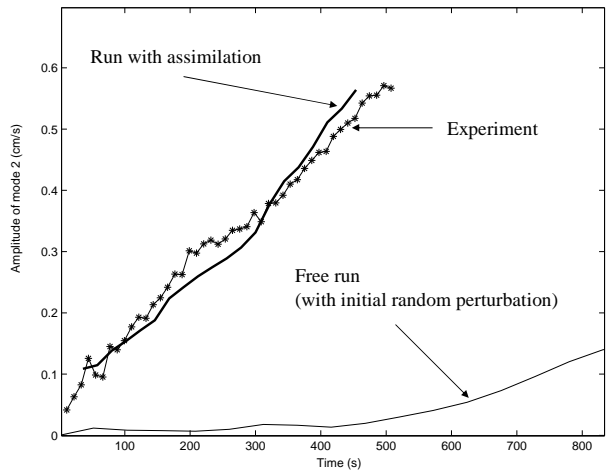}
\caption{
Amplitude of baroclinic mode 2 in the top layer as a function
of time in the experiment
(line with stars),
in the free simulation (thin line)
and in the simulation performed with data assimilation every 22s (thick line).
}
\end{figure}

\begin{figure}[p]
\centering
\includegraphics
[scale=0.8,bb=77 51 571 597,clip]
{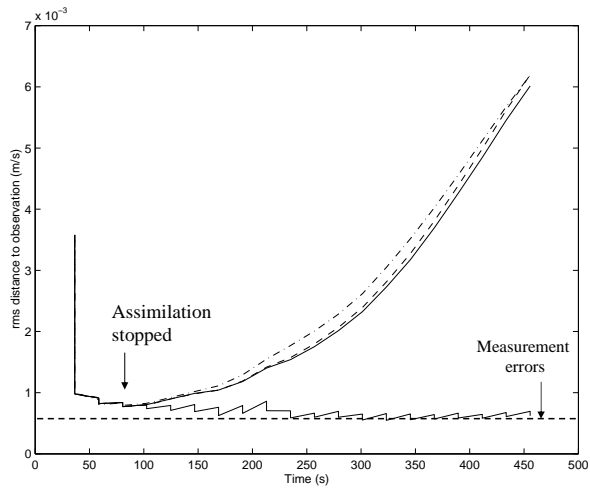}
\caption{
Value of the rms distance between the simulated and measured velocity
fields in the top layer as
a function of time in the simulation performed with data assimilation every
22s
and in the simulation where data assimilation is stopped at $t=75s$. 200$^2$ grid points
are used in both layers. The results obtained in two other test runs are also plotted :
simulation with doubled viscosity coefficient (dashed line) and simulation with additional friction (dot-dashed
line). The Ekman friction coefficient $C_f$ is taken as equal to $1.4 \; 10^{-3}s^{-1}$ in the
bottom layer and $5.6 \; 10^{-3}s^{-1}$ in the top layer. These values are those obtained
assuming rigid upper and lower boundaries.
}
\end{figure}

\begin{figure}[p]
\centering
\includegraphics
[scale=0.8,bb=77 51 571 597,clip]
{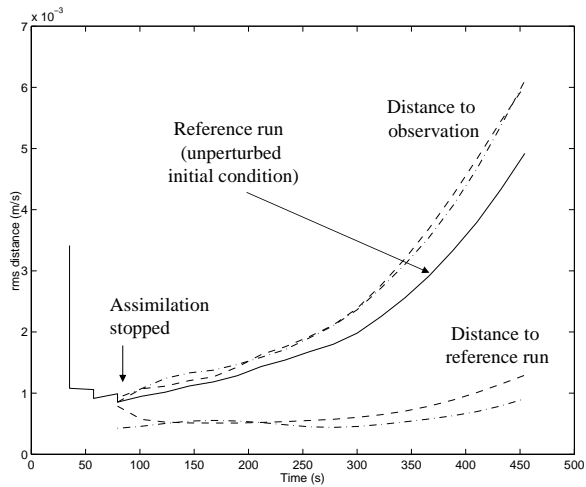}
\caption{
Value of the rms distance between the simulated and measured velocity
fields in the top layer as
a function of time when assimilation is stopped at $t=75s$.
100$^2$ grid points
are used in both layers. Two test runs were performed using a
perturbed initial condition at time $t=75s$. In the first test run (dashed line),
the velocity field $\vu$ in each layer is replaced at time
$t=75s$ by $\vu_{obs}+\bf{R} \; (\vu-\vu_{obs})$,
where $\vu_{obs}$ is the observed velocity field and $\bf{R}$ is
the 90$^o$ rotation operator.
In the second test run (dot-dashed line), a large
friction coefficient is imposed ($C_f=1.4 \; 10^{-2}s^{-1}$ in the
bottom layer and $5.6 \; 10^{-2}s^{-1}$ in the top layer) from $t=0$ to $t=75s$ only,
resulting in a slightly perturbed initial condition at time $t=75s$.
}
\end{figure}

\begin{figure}[p]
\centering
\includegraphics
[scale=0.8,bb=77 51 571 597,clip]
{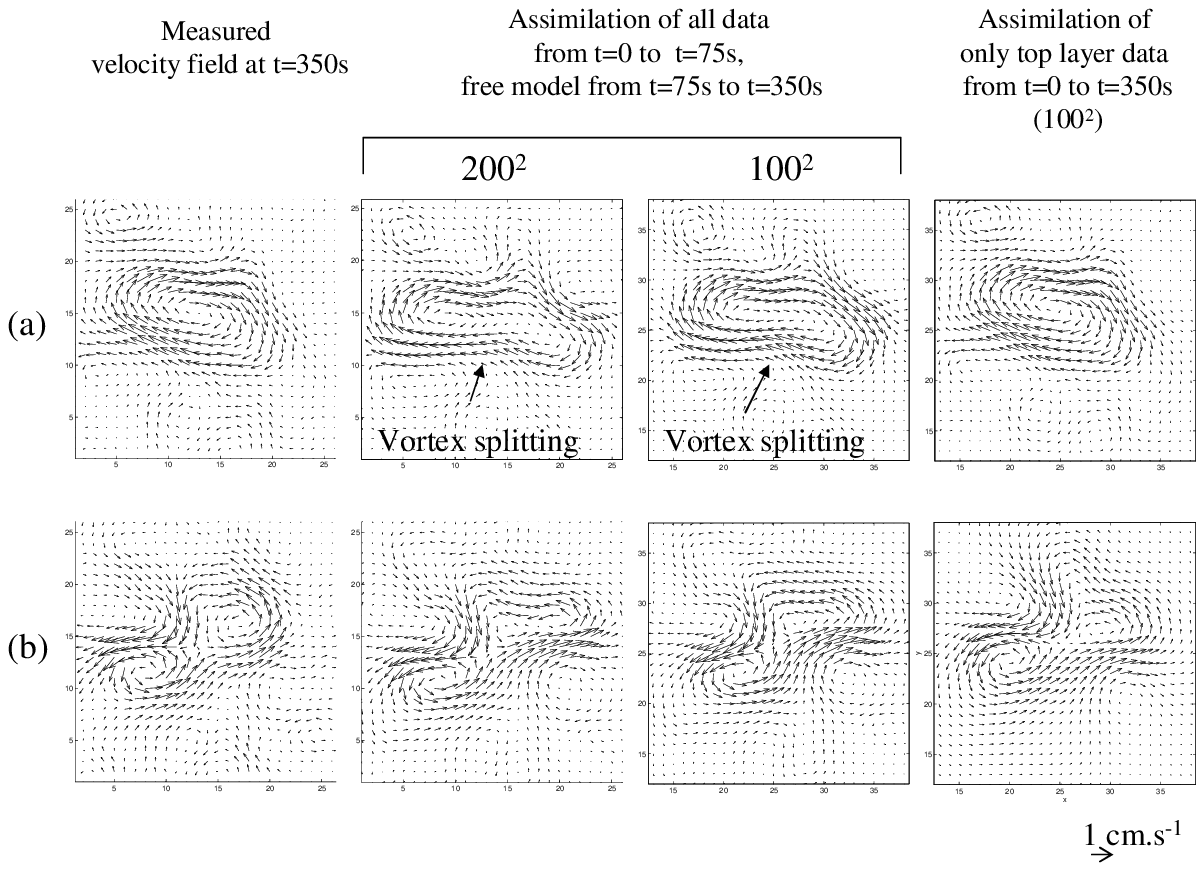}
\caption{
Velocity field in the top (a) and bottom (b) layers at $t=350s$ obtained
in the experiment and in the simulation using different assimilation
scenarios :
assimilation of all data swithched off at $t=75s$ (note that the vortex splitting occurs faster
than in the experiment, independently of the resolution);
assimilation using only top layer data until $t=350s$ (note that the bottom layer is well
reconstructed). For clarity,
only 25$^2$ vectors are plotted in all cases.
}
\end{figure}


\section{\label{conclu}Conclusion}

The results reported in the present letter 
illustrate the interest of an experimental test-bed 
for operational oceanography :
 
(i) Thanks to data assimilation, a complete description of 
the experimental flow fields is obtained, including the non-measured variables. 
Any physical 
quantity can then be calculated.  
Data assimilation can thus be used as a complementary tool for experimental investigation
and physical analysis of the flow. For instance, potential vorticity anomalies
can be calculated, providing quantities which are generally impossible to measure but which are 
crucial to a better understanding of the physics of the baroclinic instability.  

(ii) The obtained flow field can be used as an initial condition to test 
the numerical model. The divergence of the numerical model from reality is, in principle, 
caused either by the sensitivity of system evolution to the initial condition, or 
by the model error itself. We have checked that, in our test cases 
sensitivity to weak variations 
in the initial condition is not the dominant effect. 
This makes it possible to quantify 
the systematic forecast errors. Thus, even weak model errors can be detected, of the order of 
1/30 of the dominant 
inertial term in the present case. Such a weak model limitation would be probably much 
more difficult to detect in complex oceanic applications. 
Test runs were performed to show that these 
model errors are not caused by resolution, dissipation or friction problems. The most 
probable sources of error are the hydrostatic approximation or the two-layer formulation of the 
equations.  
Further work is 
needed to test this hypothesis.

(iii) The accuracy of the assimilation scheme can also be analysed in detail. 
The present study shows, for instance, how the assimilation scheme is able to reconstruct
the velocity field of the lower
layer from observation of the upper layer. This is clearly of practical 
interest because vertical extrapolation of the measured surface quantities 
is a great challenge in oceanography (see for instance Hurlburt 1986).

Many other tests can of course be performed with the available data using  
various dynamical models and/or assimilation schemes. 
Possible improvement by
non-hydrostatic models would be of particular interest. 
The study of other processes is in progress,
involving the instability of boundary currents, gravity currents on a
slope and current/topography interaction. The measurements obtained from 
these experiments are available to other researchers on the Coriolis 
web site (www.coriolis-legi.org) as a data 
base to test numerical models and assimilation schemes.  

\pagebreak

This study has been  sponsored by EPSHOM, contract
Nr. 9228. We acknowledge the kind support of Y. Morel for the implementation of
the MICOM model, 
and of J.M. Brankart,  P. Brasseur and C.E. Testut for the implementation
of the SEEK assimilation scheme.



\begin{thebibliography}{}

\bibitem{blec86}
Bleck, R. and Boudra, D. Wind driven spin-up in eddy-resolving
ocean models formulated in isopycnic coordinates,
{\it J. Geophys. Res.}, {\it 91},
7611--7621, 1986.

\medskip

\bibitem{bras99}
Brasseur, P., Ballabrera-Poy, J. \& Verron, J. 1999. Assimilation of altimetric
data in the mid-latitude oceans using the Singular Evolutive Extended
Kalman filter
with an eddy-resolving, primitive equation model, {\it J. Marine Sc.} {\it
22}, 269--294, 1999.

\medskip

\bibitem{grli81}
Fincham, A. and Delerce, G. Advanced optimization of correlation imaging
velocimetry algorithms,
{\it Experiments in Fluids} {\it  29}, S13-S22, 2000.

\medskip

\bibitem{fide00}
Griffiths, R.W. and Linden, P.F. The stability of vortices in a
rotating, stratified fluid.
{\it J. Fluid Mech.} {\it  105}, 283-316, 1981.

\medskip

\bibitem{hurl86}
Hurlburt, H.E. 
Dynamic Transfer of Simulated Altimeter Data Into
Subsurface Information by a Numerical Ocean Model. 
{\it J. Geophys. Res.} {\it  91}, C2, 2372-2400, 1986.  

\medskip

\bibitem{pham98}
Pham, D., Verron, J. and Roubaud, M. A Singular Evolutive Extended
Kalman filter for data assimilation in oceanography,
{\it J. Marine Sc.},
{\it 16} (3-4), 323--340, 1998.

\bibitem{stone71}
Stone, P.H. Baroclinic instability under non-hydrostatic conditions. 
{\it J. Fluid Mech.} {\it  45}, part 4, 659-671, 1971.

%

\end{thebibliography}
\end{document}